\documentclass[aip,rsi,amsmath,amssymb,reprint,]{revtex4-1}

\usepackage{graphicx}% Include figure files
\usepackage{dcolumn}% Align table columns on decimal point
\usepackage{bm}% bold math
\begin{document}

\title{Miniaturized high-frequency sine wave gating InGaAs/InP single-photon detector}

\author{Wen-Hao Jiang}
\affiliation{Hefei National Laboratory for Physical Sciences at Microscale and Department of Modern Physics,
University of Science and Technology of China, Hefei, Anhui 230026, China}
\affiliation{CAS Center for Excellence in Quantum Information and Quantum Physics,
University of Science and Technology of China, Hefei, Anhui 230026, China}

\author{Xin-Jiang Gao}
\email{gaoxinjiang@coeri.com}
\affiliation{China Electronics Technology Group Corporation No.44 Research Institute, Chongqing 400060, China}

\author{Yu-Qiang Fang}
\affiliation{Hefei National Laboratory for Physical Sciences at Microscale and Department of Modern Physics,
University of Science and Technology of China, Hefei, Anhui 230026, China}
\affiliation{CAS Center for Excellence in Quantum Information and Quantum Physics,
 University of Science and Technology of China, Hefei, Anhui 230026, China}

\author{Jian-Hong Liu}
\affiliation{Quantum CTek Co., Ltd., Hefei, Anhui 230088, China}

\author{Yong Zhou}
\affiliation{China Electronics Technology Group Corporation No.44 Research Institute, Chongqing 400060, China}

\author{Li-Qun Jiang}
\affiliation{China Electronics Technology Group Corporation No.44 Research Institute, Chongqing 400060, China}

\author{Wei Chen}
\affiliation{China Electronics Technology Group Corporation No.44 Research Institute, Chongqing 400060, China}

\author{Ge Jin}
\affiliation{Hefei National Laboratory for Physical Sciences at Microscale and Department of Modern Physics,
University of Science and Technology of China, Hefei, Anhui 230026, China}
\affiliation{CAS Center for Excellence in Quantum Information and Quantum Physics,
 University of Science and Technology of China, Hefei, Anhui 230026, China}

\author{Jun Zhang}
\email{zhangjun@ustc.edu.cn}
\affiliation{Hefei National Laboratory for Physical Sciences at Microscale and Department of Modern Physics,
University of Science and Technology of China, Hefei, Anhui 230026, China}
\affiliation{CAS Center for Excellence in Quantum Information and Quantum Physics,
University of Science and Technology of China, Hefei, Anhui 230026, China}

\author{Jian-Wei Pan}
\affiliation{Hefei National Laboratory for Physical Sciences at Microscale and Department of Modern Physics,
University of Science and Technology of China, Hefei, Anhui 230026, China}
\affiliation{CAS Center for Excellence in Quantum Information and Quantum Physics,
University of Science and Technology of China, Hefei, Anhui 230026, China}

\date{\today}

\begin{abstract}
High-frequency gating InGaAs/InP single-photon detectors (SPDs) are widely used for applications requiring single-photon detection in the near-infrared region such as quantum key distribution. Reducing SPD size is highly desired for practical use, which is favorable to the implementation of further system integration.
Here we present, to the best of our knowledge, the most compact high-frequency sine wave gating (SWG) InGaAs/InP SPD. We design and fabricate an InGaAs/InP single-photon avalanche diode (SPAD) with optimized semiconductor structure, and then encapsulate the SPAD chip and a mini-thermoelectric cooler inside a butterfly package with a size of 12.5 mm $\times$ 22 mm $\times$ 10 mm. Moreover, we implement a monolithic readout circuit for the SWG SPD in order to replace the quenching electronics that is previously designed with board-level integration. Finally, the components of SPAD, monolithic readout circuit and the affiliated circuits are integrated into a single module with a size of 13 cm $\times$ 8 cm $\times$ 4 cm. Compared with the 1.25 GHz SWG InGaAs/InP SPD module (25 cm $\times$ 10 cm $\times$ 33 cm) designed in 2012, the volume of our miniaturized SPD is reduced by 95\%. After the characterization, the SPD exhibits excellent performance with a photon detection efficiency of 30\%, a dark count rate of 2.0 kcps and an afterpulse probability of 8.8\% under the conditions of 1.25 GHz gating rate, 100 ns hold-off time and 243 K. Also, we perform the stability test over one week, and the results show the high reliability of the miniaturized SPD module.
\end{abstract}

%%%%%%%%%%%%%%%%%%%%%%%%%%  body  %%%%%%%%%%%%%%%%%%%%%%%%%%
\maketitle

\section{Introduction}
Single-photon detectors (SPDs) have the ultimate sensitivity for weak light detection, and thus are widely used in numerous applications such as quantum key distribution (QKD), lidar and photoluminescence. Currently, the primary SPD technologies in the near-infrared region include superconducting nanowire single-photon detector (SNSPD)~\cite{FVJ13,ZYL17}, up-conversion single-photon detector~\cite{SPW13} and InGaAs/InP single-photon avalanche diode (SPAD)~\cite{ZIZ15}, among which InGaAs/InP SPAD is preferred in practical applications due to the advantages of small size and low cost. In InGaAs/InP SPDs, SPADs are operated either in free-running mode or gating mode. Free-running mode is a natural method for single-photon detection, and particularly suitable for the applications with unknown photon arrival times. So far, diverse techniques have been reported to implement free-running InGaAs/InP SPDs~\cite{RWR00,LHC08,ZTG09,WIB09,IJN09,YHH12,KWL14,YS17}. However, the SPD count rates in these schemes are considerably limited, due to the fact that long hold-off time has to be applied to suppress the afterpulsing effect.

High-frequency gating technique, including sine wave gating (SWG)~\cite{SWG06} and self-differencing~\cite{SD07}, provides a practical approach to significantly increase SPD count rate, which is highly desired in applications requiring high detection rate. For instance, in QKD applications secure key rate is roughly proportional to system clock rate, so that using high-frequency gating InGaAs/InP SPD can substantially improve QKD performance. In the high-frequency gating scheme, gating rate usually reaches
a gigahertz level and the avalanche duration time is limited to a few hundred picoseconds, which effectively suppresses the charge carrier quantity of avalanche and thus, the afterpulse probability. As a consequence of ultrashort gating time, the avalanche signals become pretty weak, i.e., at the level of mV. Therefore,
the key challenge in high-frequency gating scheme is to extract weak avalanche signals from the large capacitive response signals~\cite{ZIZ15}.

Concretely, in the SWG scheme, sine waves with large amplitude are gated on an InGaAs/InP SPAD. Due to the pure and simple frequency spectrum of sine waves, the capacitive response signals from the SPAD only include the fundamental frequency and higher-order harmonics of sine waves, which can be easily eliminated by filters. After filtering, the weak avalanche signals are amplified and then discriminated. So far, many groups have reported the implementations of SWG schemes with InGaAs/InP SPADs~\cite{SWG06,NAI09,ZTB09,GAP10,RGL11,NEC11,LLW12,WLG12,NIST13,HWC17,JLL17} and even silicon SPADs~\cite{SNT17,ZJC17}.
For practical applications, integration implementation of SPD is crucial. For instance, in 2012 Liang \emph{et al.} reported a board-level integrated SWG InGaAs/InP SPD with a gating rate of 1.25 GHz and a size of 25 cm $\times$ 10 cm $\times$ 33 cm~\cite{LLW12},
in which a commercial InGaAs/InP SPAD with transistor outline package was employed such that a bulky metal box with a thermoelectric cooler inside had to be used for cooling and numerous discrete electronic components were exploited to design the quenching electronics.
Subsequently, such SPD modules had been further integrated into high-speed QKD system~\cite{MWZ18}, in which the size ratio of SPD modules to the QKD receiver system reached $\sim$ 60\%. This indicates that the integration level of current high-frequency SWG InGaAs/InP SPD has to be extremely improved in order to meet the requirements of the next-generation miniaturized QKD system.

In this paper, we present,
to the best of our knowledge, the world's smallest SWG InGaAs/InP SPD with 1.25 GHz gating rate and 13 cm $\times$ 8 cm $\times$ 4 cm in size. Compared with the board-level integrated SWG SPD~\cite{LLW12}, the SPD module size has been drastically reduced by 95\%.
On one hand, we design and fabricate a high-performance InGaAs/InP SPAD, and further integrate the SPAD chip and a mini-thermoelectric cooler into a single butterfly package component with a size of 12.5 mm $\times$ 22 mm $\times$ 10 mm. On the other hand,
we implement a monolithically integrated readout circuit (MIRC) with 15 mm $\times$ 15 mm in size to replace the conventional quenching electronics designed with board-level integration. Our miniaturized
SWG SPD exhibits excellent performance with 30\% photon detection efficiency (PDE), 2.0 kcps dark count rate (DCR) and 8.8\% afterpulse probability ($P_{ap}$) at 243 K and 100 ns hold-off time.

\section{Miniaturized SPD design}
\subsection{Design and fabrication of InGaAs/InP SPAD}

\begin{figure}[tbp]
\centering
\includegraphics[width=8 cm]{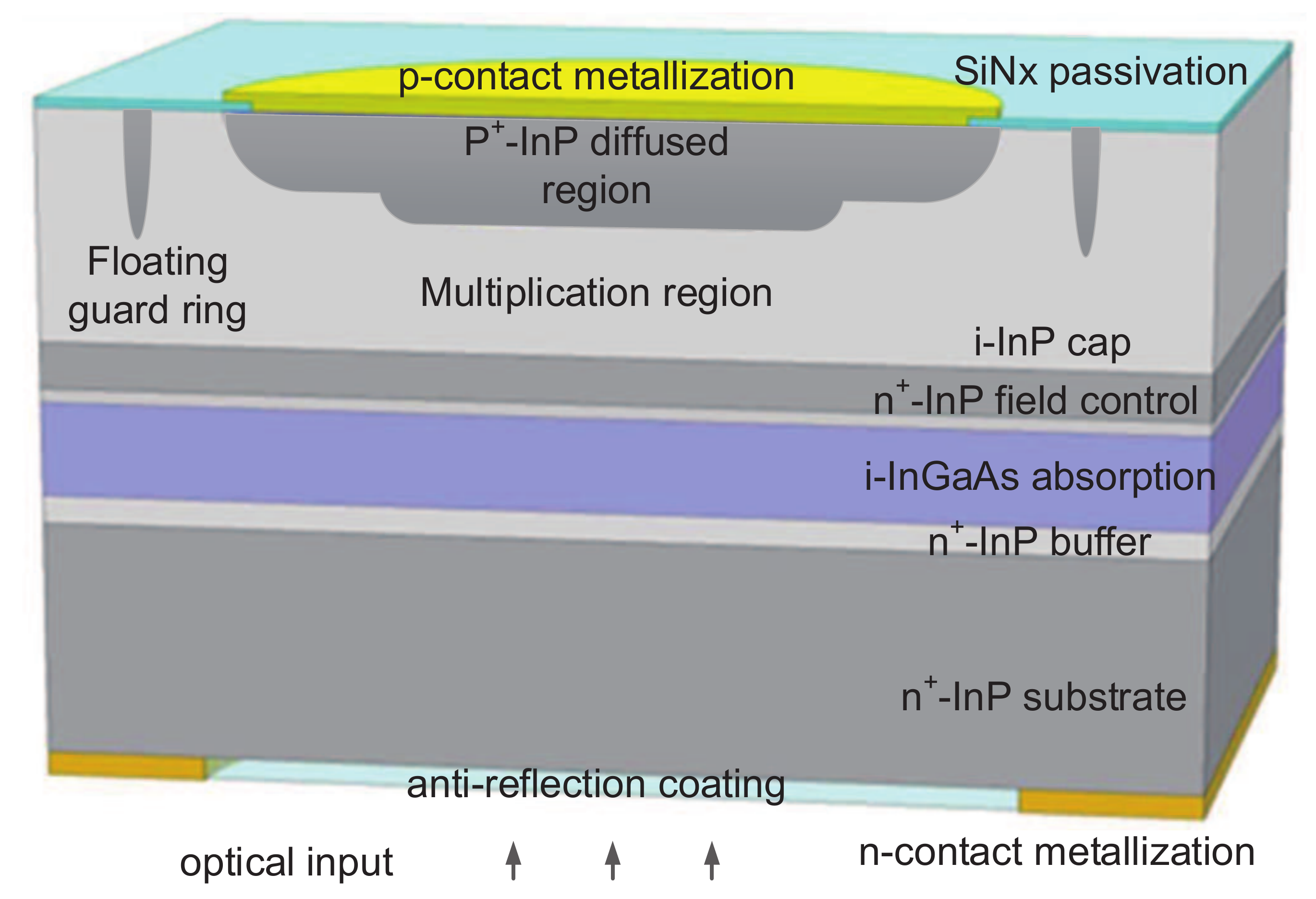}\\
\caption{Semiconductor structure diagram for high-performance InGaAs/InP SPAD design.}
\label{fig1}
\end{figure}

We follow the method~\cite{MBW16} to design an InGaAs/InP SPAD with separate absorption, grading, charge, and multiplication (SAGCM) structure~\cite{ZIZ15,IBH07}, as illustrated in Fig.~\ref{fig1}. Further, we combine the semiconductor structure optimization, particularly for InGaAs absorption layer, InP electric field control layer and InP multiplication layer, and fabrication process control to improve the performance of key SPAD parameters including PDE, DCR, afterpulse probability and timing jitter.
By tuning the doping concentration in the electric field control layer, the electric field distributions in the InGaAs absorption layer and the InP multiplication layer are well controlled, which can also reduce the timing jitter of the SPAD chip.

In the absorption layer, on one hand the electric field strength should be as low as to avoid avalanche breakdown, and on the other hand the electric field strength should be as high as to maintain the saturation drift rate of the photogenerated hole carriers. Then, the electric field strength in this layer is compromised at the range of 1 $\sim$ 1.4 $\times 10^5$ V/cm.
In the multiplication layer, properly decreasing the thickness can reduce the transit time of carriers and thus the timing jitter whilst maintain the enough electric field strength for avalanche breakdown. The multiplication layer thickness is controlled at the range of 1 $\sim$ 1.5 $\mu$m.
In the PN junction, a ladder structure is designed, which can be formed using double-diffusion process technology. Such structure has several advantages, i.e., reshaping the electric field distribution of the junction, suppressing the marginal effect of electric field due to curvature, creating uniform electric field in the central zone of the PN junction, and reducing the high electric field regions at the edge of the PN junction and thus the afterpulse probability.
Moreover, a floating guard ring is designed to absorb carriers outside the active region, which can help reduce DCR and improve the response time of SPAD chip.

The SPAD chip is fabricated using the standard epitaxial process of metal-organic chemical vapour deposition (MOCVD). During the process, the background impurity concentration of the epitaxial material reaches as low as 2 $\times 10^{14}$/cm$^3$, whilst the error of surface charge density in the electric field control layer is controlled at 5\%. After the diffusion with a Zn source,
the error of depth difference between the first and the second P-doping processes is controlled at 50 nm, and finally a central PN junction with a diameter of 25 $\mu$m is formed in the multiplication layer.

\begin{figure}[tbp]
\centering
\includegraphics[width=9 cm]{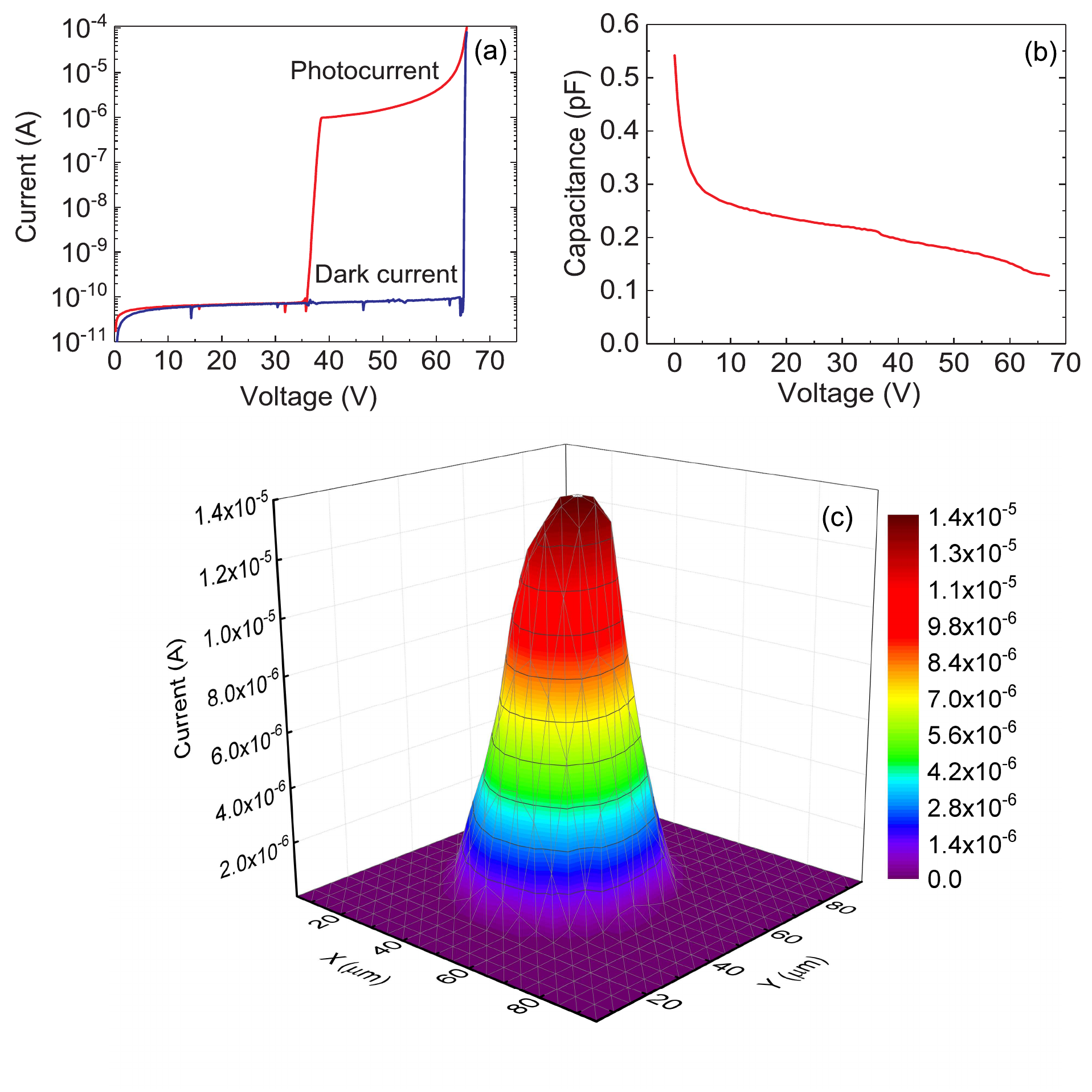}\\
\caption{Measured I-V curve (a), C-V curve (b), and two-dimensional photocurrent distribution in the linear mode (c) of SPAD.}
\label{fig2}
\end{figure}

We then characterize the chip before packaging at room temperature, as shown in Fig.~\ref{fig2}. Fig.~\ref{fig2}(a) plots the current-voltage (I-V) curves with and without light illumination, from which one can observe that the punch-through and breakdown ($V_{br}$) voltages of SPAD chip are 38.6 V and 65.9 V, respectively. With a reverse bias voltage of $V_{br}-$1 V, the total dark current of SPAD chip is as low as 0.2 nA, which is primarily contributed by the surface dark current. Fig.~\ref{fig2}(b) plots the capacitance-voltage (C-V) curve, in which the capacitance of SPAD chip decreases rapidly with bias voltage less than $\sim$ 5 V and then decreases slowly as bias voltage increases. When the SPAD is operated in Geiger mode, the capacitance of SPAD chip is $\sim$ 0.13 pF.

In order to further calibrate the light response of SPAD chip, we perform a full two-dimensional photocurrent-position scan using a point light source
of 5 $\mu$m spot size with a position step of 2 $\mu$m and a bias voltage of $V_{br}-$1 V, as shown in Fig.~\ref{fig2}(c).
When the light source is located to the center of active region, the photocurrent generated by the SPAD chip is largest. As the light source moves towards the edge of active region, the photocurrent drastically decreases. When the light source illuminates on the N-contact metal layer outside the active region, the photocurrent approximates to zero. This indicates that the double-diffusion process successfully realizes high electric field in the central region of the PN junction whilst effectively suppresses the electric field of the marginal region in the multiplication layer.

After the calibration, the SPAD chip is encapsulated inside a butterfly package (12.5 mm $\times$ 22 mm $\times$ 10 mm) along with a mini-thermoelectric cooler (TEC). The SPAD chip is fixed on the cold side of TEC and then fiber pigtailed. The hot side of TEC is soldered on the bottom of the metal box for heat dissipation. The operation temperature of SPAD chip is monitored via a negative temperature coefficient thermistor very close to the chip.

\subsection{MIRC}

In the readout circuit of SWG scheme, either band-stop filters~\cite{SWG06,NAI09,GAP10,NEC11,LLW12,JLL17} or low-pass filters (LPFs)~\cite{WLG12,JLL17} are used to eliminate the capacitive response signals, and a low-noise amplifier (LNA) is used to amplify the avalanche signals. Here we implement a monolithic readout circuit with a size of 15 mm $\times$ 15 mm, and the details of the MIRC can be found in Ref.~\cite{JLL17}.
Two LPFs and a two-stage LNA are integrated inside the MIRC chip, which is fabricated using the technology of low temperature co-fired ceramics (LTCC).

LTCC is a standard integration and fabrication technique for radio frequency miniaturized components. During the LTCC process, via holes are punched into ceramic tapes and filled with silver paste for electrical connections between layers, and the circuits are printed onto the tapes. Then the tapes are stacked, laminated and cut in a sequential order, to form a desired shape. Finally, the silver paste and ceramic tapes are co-fired together at 900 $^\circ$C.

The MIRC chip is then tested using a network analyzer, and exhibits excellent ratio frequency performance with a gain of $\sim$ 40 dB below 1 GHz and a rejection ratio of $\sim$ 80 dB at 1.25 GHz. This result indicates that the MIRC chip can effectively filter out the capacitive response signals in the SWG scheme with a gating frequency of
1.25 GHz and extract weak avalanche signals.

\subsection{SPD Module}

\begin{figure}[tbp]
\centering
\includegraphics[width=8 cm]{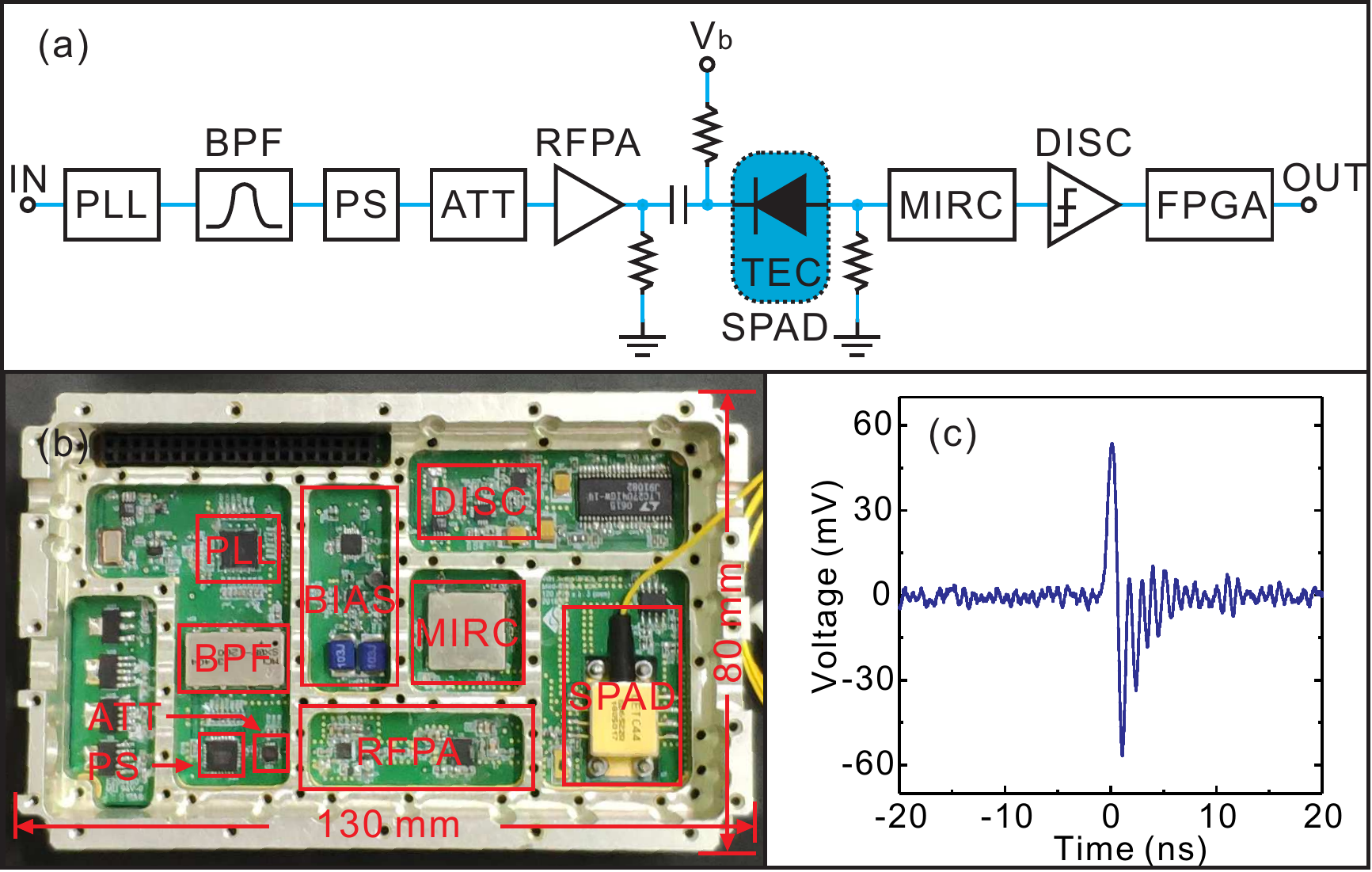}\\
\caption{The design diagram (a) and photo (b) of the miniaturized 1.25 GHz InGaAs/InP SWG SPD. (c) Typical avalanche signal at the output of MIRC captured by an oscilloscope with 8 GHz bandwidth.}
\label{fig3}
\end{figure}

We then integrate the SPAD component, MIRC and the affiliated circuits together to implement a miniaturized 1.25 GHz InGaAs/InP SWG SPD with a size of 13 cm $\times$ 8 cm $\times$ 4 cm. To the best of our knowledge, such SPD module is the most compact so far. Fig.~\ref{fig3}(a) and Fig.~\ref{fig3}(b) show the design diagram and the photo of SPD, respectively. An external 10 MHz clock is used as the reference signals of phase-locked loop (PLL). The PLL generates initial 1.25 GHz square wave signals. A band-pass filter (BPF) converts the square wave signals into sine wave gates. The amplitude and phase of sine wave gates are regulated by an attenuator (ATT) and a six-bit phase shifter (PS), respectively. After passing through a narrow-band radio frequency power amplifier (RFPA), the sine wave gates are amplified up to 10 V of $V_{pp}$, and then alternating current (AC) coupled to the cathode of SPAD.

At the anode of SPAD, the MIRC chip directly extracts weak avalanche signals via a sample resistor of 50 $\Omega$. Fig.~\ref{fig3}(c) depicts a typical avalanche trace captured by a high-speed oscilloscope. Then the avalanche signals are discriminated by a high-speed discriminator (DISC) and the output signals are entered into a field programmable gate array (FPGA), in which hold-off time setting is applied to further suppress the afterpulsing effect.
Hold-off time means that once an avalanche is triggered the avalanches during the following time period would not be counted~\cite{ZIZ15,ZTB09}.
Also, FPGA is used to control the parameters of ATT, PS and TEC inside the SPAD component. The SPD module is supplied by a single 12 V direct current (DC) source with a total power dissipation of 15 W.

\section{SPD characterization}

The miniaturized SWG SPD module is characterized using the standard calibration method~\cite{ZIZ15}. A signal generator outputs synchronized signals, including 10 MHz signal for the reference clock of SPD and 625 kHz signal for triggering the pulsed laser and the time-to-digital converter (TDC).
The laser pulses with a width of $\sim$ 100 ps are divided by a 99:1 beam splitter. The 99\% port is monitored by a power meter in real-time, and the pulses from the 1\% port enter into an optical variable attenuator to further attenuate the intensity down to a level of mean photon number per pulse of 1. The detection outputs of SPD are used as ``stop'' signals of TDC, from which the key parameters of PDE, DCR and $P_{ap}$ can be measured.

\begin{figure}[tbp]
\centering
\includegraphics[width=8.5 cm]{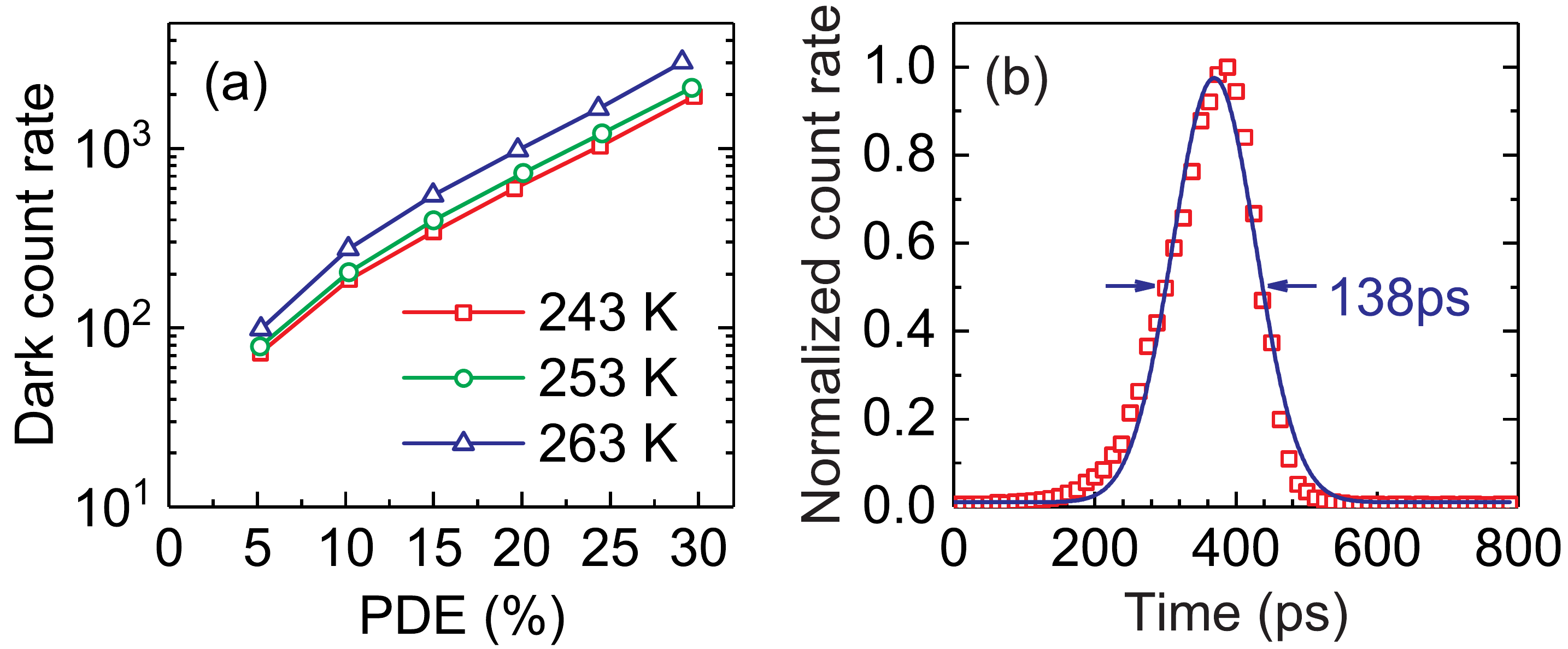}\\
\caption{(a) DCR versus PDE with three temperatures. (b) Effective gating width measurement at 20\% PDE and 243 K.}
\label{fig4}
\end{figure}

PDE and DCR are intrinsic parameters of SPAD, which are independent from the readout circuits. Fig.~\ref{fig4}(a) plots DCR as a function of PDE at 243 K, 253 K and 263 K, respectively. For a Poissonian light source, PDE is calculated by~\cite{ZIZ15}
\begin{equation}
\label{PDE}
PDE=\frac{1}{\mu}\ln\frac{1-DCR/f_{g}}{1-R_{ph}/f_{l}},
\end{equation}
where $\mu$ is the mean photon number per laser pulse, DCR is the measured count rate without laser illumination, $f_{g}$ is the gating frequency, $f_{l}$ is the frequency of laser pulses, and $R_{ph}$ is the photon detection count rate with laser illumination, i.e., the coincidence rate between detections and laser pulses with the subtraction of DCR contribution. As plotted in Fig.~\ref{fig4}(a), the SPD exhibits excellent performance, e.g., at PDE of 30\% DCR reaches as low as 2.0 kcps.

Effective gating width is an important parameter related to charge carrier quantity of avalanche, which is measured by scanning the relative delay between laser pulses and sine wave gates. For instance, as fitted in Fig.~\ref{fig4}(b), the effective gating width is 138 ps full width at half maximum (FWHM) at 20\% PDE and 243 K. Such low value can significantly suppress the afterpulsing effect~\cite{ZIZ15}.

\begin{figure}[bp]
\centering
\includegraphics[width=8.5 cm]{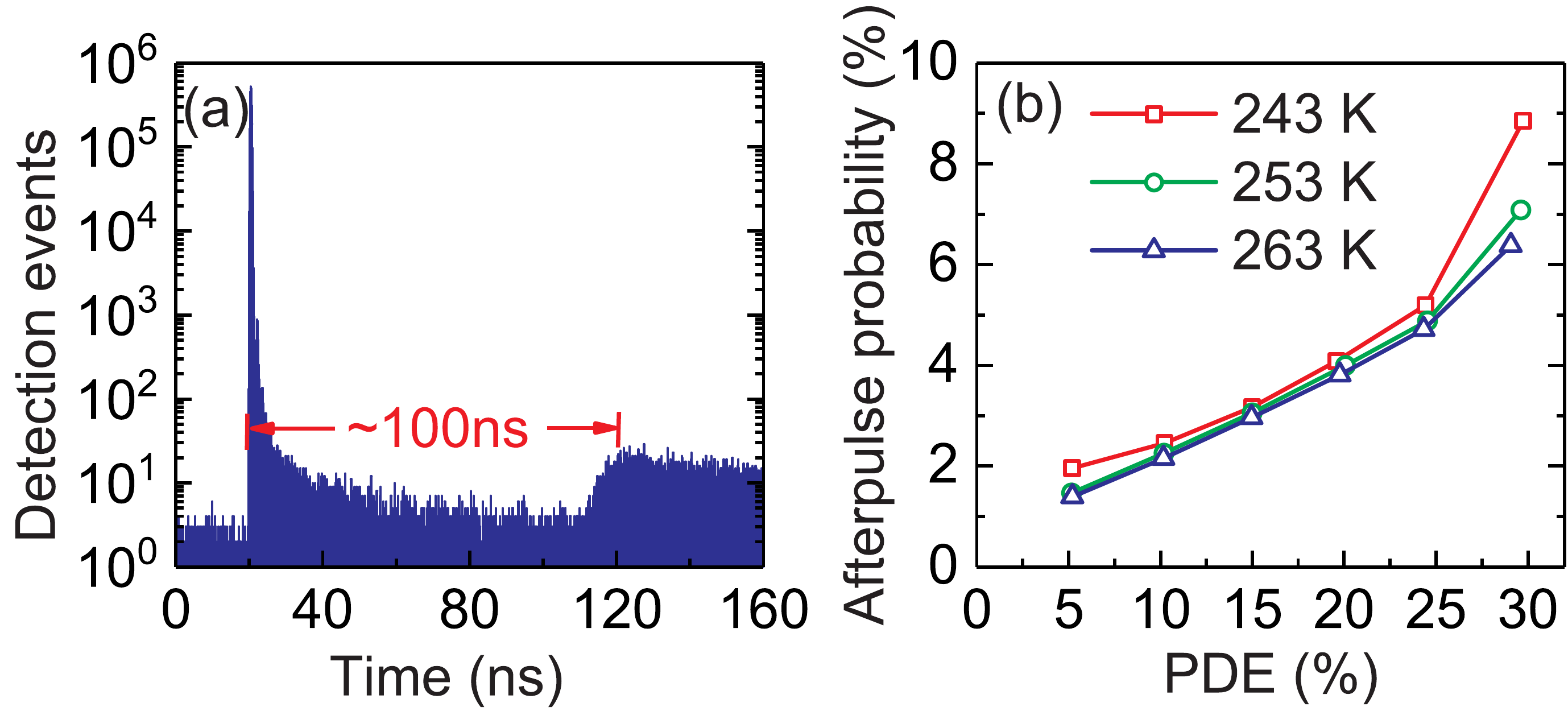}\\
\caption{(a) The histogram of detection event distribution under the conditions of 20\% PDE, 243 K and 100 ns hold-off time. (b) $P_{ap}$ versus PDE with three temperatures.}
\label{fig5}
\end{figure}

Apart from DCR, the afterpulse probability is another noise source of SPD that is related to other parameters and readout circuits. During the characterization, with the pulsed laser illumination the detection event distribution is first recorded by the TDC. Fig.~\ref{fig5}(a) illustrates a typical histogram of detection event distribution
under the conditions of 20\% PDE, 243 K and 100 ns hold-off time.
The main peak corresponds to the photon detection counts, and the subsequent decay of detection events is attributed to afterpulse counts. The sudden increase of detection events at 120 ns in Fig.~\ref{fig5}(a) is due to the effect of hold-off time. From the detection event distribution $P_{ap}$ is calculated after subtracting the DCR contribution~\cite{LLW12}.

Fig.~\ref{fig5}(b) plots $P_{ap}$ as a function of PDE.
At 243 K and 30\% PDE, $P_{ap}$ is 8.8\%. As temperature increases to 253 K and 263 K, with the same PDE $P_{ap}$ decreases to 7.1\% and 6.4\%, respectively. One can further normalize the $P_{ap}$ values for comparison.
For instance, at 243 K and 30\% PDE, $P_{ap}$/ns is 8.8\% /(5.3 $\mu$s $\times$ 176 ps/800 ps) $\sim 7.5 \times 10^{-5}$/ns, where 5.3 $\mu$s is the average time interval between photon detection events and 176 ps is the measured effective gating width. Such normalized value is around 8 times higher than the normalized DCR parameter, i.e., 2.0 kcps/(176 ps/800 ps) $\sim 9.1\times 10^{-6}$/ns.

\begin{figure}[tbp]
\centering
\includegraphics[width=8.5 cm]{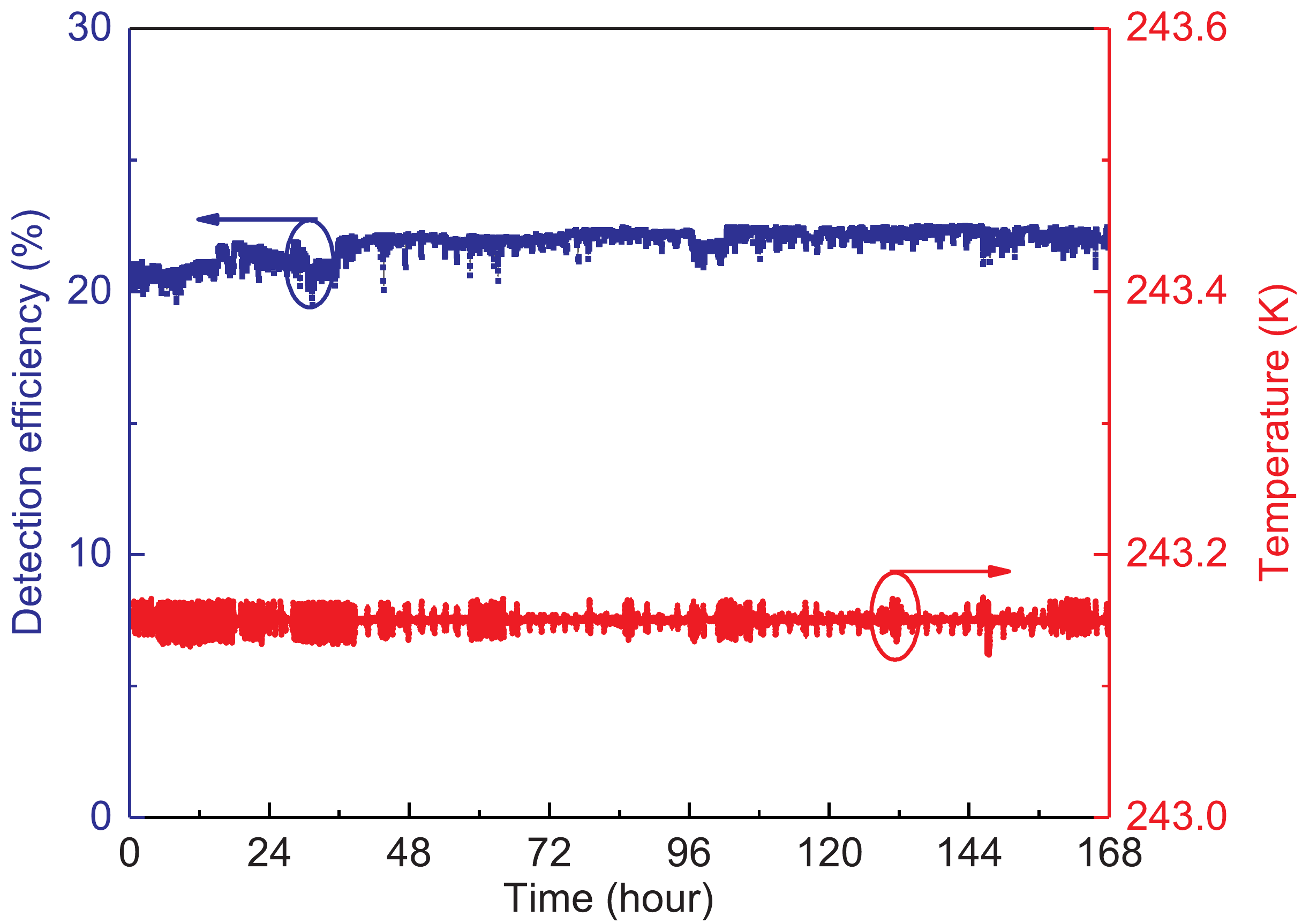}\\
\caption{Stability test of the miniaturized SWG SPD module.}
\label{fig6}
\end{figure}

Finally, we perform the stability test over one week for the miniaturized SWG SPD module by monitoring two key parameters including PDE and temperature.
Fig.~\ref{fig6} depicts the test results. During the test, the SPD module is connected with a computer via RS-232 serial port. PDE is initially set to $\sim$ 20\%.
The values of detection count and temperature are recorded every second. Every 5 minutes, the SPD runs a full scan for the PS during 30 seconds to optimize the relative delay between sine wave gates and laser pulses. The two straight lines in Fig.~\ref{fig6} clearly indicate the stability of the miniaturized SWG SPD module, which can be ready for practical use.

\section{Conclusion}

In conclusion, we have reported, to the best of our knowledge, the most compact SWG InGaAs/InP SPD with 1.25 GHz gating frequency. We have designed and fabricated a high-performance InGaAs/InP SPAD component integrated with a mini-thermoelectric cooler. Further, we have implemented a monolithic readout circuit for the SWG scheme.
With the help of the miniaturized SPAD component and the monolithic readout circuit, the SPD module has been integrated within a size of
13 cm $\times$ 8 cm $\times$ 4 cm. Compared with the board-level integrated SPD designed in 2012~\cite{LLW12}, such size has been reduced by 95\%.
The SPD exhibits excellent performance with 30\% PDE, 2.0 kcps DCR and 8.8\% afterpulse probability at 243 K and 100 ns hold-off time, and stability test results show that the miniaturized SPD module can be used in practical applications. Our work paves the way to implement the miniaturization of high clock rate QKD system in the future.

\section*{acknowledgements}

This work has been supported by the National Key R\&D Program of China under Grant No.~2017YFA0304004, the National Natural Science Foundation of China under Grant No.~11674307, the Chinese Academy of Sciences, and the Anhui Initiative in Quantum Information Technologies.

%\section*{Acknowledgments}
%Acknowledgments, if included, should appear at the end of the document. The section title should not be numbered.

\end{document}